\documentclass[a4paper,twoside]{article}

\usepackage{epsfig}
\usepackage{subcaption}
\usepackage{calc}
\usepackage{amssymb}
\usepackage{amstext}
\usepackage{amsmath}
\usepackage{amsthm}
\usepackage{multicol}
\usepackage{apalike}
\usepackage{algorithm2e}
\usepackage[bottom]{footmisc}
\usepackage{SCITEPRESS}   
\usepackage{hyperref} 

\begin{document}
\title{Creating a Trajectory for Code Writing: Algorithmic Reasoning Tasks}

\author{\authorname{Shruthi Ravikumar, Margaret Hamilton\thanks{https://orcid.org/0000-0002-3488-4524}, Charles Thevathayan\thanks{https://orcid.org/0000-0003-2605-1722}, Maria Spichkova\thanks{https://orcid.org/0000-0001-6882-1444}, Kashif Ali, Gayan Wijesinghe} 
\affiliation{
School of Computing Technologies,
RMIT University, Melbourne Victoria, Australia
}
\email{~s3613612@student.rmit.edu.au, 
\{margaret.hamilton,maria.spichkova,charles.thevathayan,kashif.ali,gayan.wijesinghe\}@rmit.edu.au 
}
}


\abstract{
Many students in introductory programming courses fare poorly in the code writing tasks of the final summative assessment. Such tasks are designed to assess whether novices have developed the analytical skills to translate from the given problem domain to coding. In the past researchers have used instruments such as code-explain and found that the extent of cognitive depth reached in these tasks correlated well with code writing ability. However, the need for manual marking and personalized interviews used for identifying cognitive difficulties limited the study to a small group of stragglers. To extend this work to larger groups, we have devised several question types with varying cognitive demands collectively called Algorithmic Reasoning Tasks (ARTs), which do not require manual marking. These tasks require levels of reasoning which can define a learning trajectory. \\
This paper describes these instruments and the machine learning models used for validating them. We have used the data collected in an introductory programming course in the penultimate week of the semester which required attempting ART type instruments and code writing. Our preliminary research suggests ART type instruments can be combined with specific machine learning models to act as an effective learning trajectory and early prediction of code-writing skills.
\\
~\\
\emph{Preprint. Accepted to the 19th International Conference on Evaluation of Novel Approaches to Software
Engineering (ENASE 2024). Final version to be published by SCITEPRESS, http://www.scitepress.org}
}

\onecolumn \maketitle \normalsize \setcounter{footnote}{0} 

\section{\uppercase{Introduction}}
\label{sec:intro}
Code writing requires students to develop solutions for the given programming problems. Students are expected to combine the various constructs, while reasoning about the overall behaviour of the resulting code.
A literature review on topics relating to novice teaching and learning has been introduced in \cite{robins2003learning}. 
Many case studies demonstrated that novices who lack reasoning skills struggle to write code, see for example \cite{denny2008evaluating}, \cite{lister2009further}, \cite{lister2006not}, and \cite{malik2019learning}.  
Thus, to help students succeed in learning programming skills, we have to, first of all, 
support them in developing reasoning skills. 
 
To determine predecessor skills required for code writing, 
so-called ``explain in plain English'' (EiPE) questions have been introduced in~\cite{lopez2008relationships}. These questions are 
classified at the relational level of the \emph{Structure of the Observed Learning Taxonomy} (SOLO) taxonomy, which can be used to measure and quantify reasoning skills. The SOLO taxonomy classifies learning and assessment tasks based on hierarchical cognitive levels, see~\cite{biggs1982system}. 
The work of Lopez et al. was an experiment designed to challenge the results presented in~\cite{mccracken2001multi}. 
McCracken et al. conducted a
 study to assess the programming ability of 216 students from four different universities. They claimed that the students' performances were poor because of their weak capacity in problem solving. The results of their study also demonstrate that when teaching  novice programmers (who may be weak in problem solving skills), an educator should consider assessing the students'  precursor skills, which are their code reading skills. Lopez et al.  used the students' performances on the EiPE question to assess the students' reasoning skills. The study found a significant relationship (a Pearson correlation of 0.5586) between code reading and code writing.

The Neo-Piagetian theory \cite{teague2013qualitative} also suggested that the novice programmers progress through various stages of learning and they can understand the overall purpose of the code once they develop abstract reasoning skills. 
Students might find problem solving difficult, because it requires combining high level thinking, problem abstraction and algorithm development, with language syntax and code tracing.  

There has been extensive exploration of approaches to teach programming, comparing languages, intelligent tutoring systems, pedagogical strategies, and conceptual methodologies, software engineering concepts, etc., 
see \cite{echeverria2017using}, \cite{silva2016more}, \cite{spichkova2019industry},  \cite{xie2019theory},  \cite{thevathayan2017combining}, 
\cite{spichkova2022teaching}, \cite{young2021project},  \cite{spichkova2018autonomous}, \cite{alharthi2018gender}, \cite{spichkova2016visual}, \cite{simic2016enhancing},  \cite{spichkova2016visual}, and \cite{pears2007survey}.   
Despite these advancements, a recent survey across 161 institutions worldwide revealed the failure rate in the introductory programming courses to be $28\%$, see~\cite{bennedsen2019failure}.  
While Bennedsen and Caspersen see the state of the art as an improvement with respect to the situation in 2007 (failure rate of $33\%$), they perceive it not to be not as alarmingly high when compared with $42-50\%$ failure rates for college
algebra in the US.  However, we perceive these numbers to be relatively high as they mean that almost every third student fails the course. Moreover, in many universities, a course with a failure rate of $20\%$ or more is flagged as requiring additional attention to the students' performance, and programming courses very often fall into this cohort. 

Thus, to have an approach to improve students' performance (without grade inflation) would be really helpful. For example, a scientific approach of learning trajectories in the field of mathematics has been shown to improve students performance, see~\cite{izu2019fostering}. Similarly, it is essential to create a trajectory between code tracing and code writing to improve students code writing abilities. 
Izy et al.  defined tentative theoretical learning trajectories that can guide teachers as they select and sequence learning activities in their introductory courses. 
In our paper, we aim to create a learning trajectory for code writing based on algorithmic reasoning.  

\emph{Contributions:}
In this study we have developed a combined approach to predict students' performance based on abstract reasoning, with the goal being to improve the students programming skills. The main contribution of our work is the \emph{Algorithmic Reasoning Tasks (ART)} framework to assess the students reasoning skills. The ART framework includes three different types of questions:

\begin{itemize} 

   \item \emph{ART Detection Type questions}, which require the in-depth study of an algorithm to determine its overall effect;

  \item \emph{ART Comparison Type questions}, which require identifying different algorithms producing the same effect;

  \item \emph{ART Algorithm analysis type questions}, which require reasoning about behaviour for specified criteria such as performance.

\end{itemize}
The framework allows automatic prediction of student performance on code writing, based on their performance on ART type questions.

\section{BACKGROUND AND RELATED WORKS}
\label{sec:background}

In this section, we discuss related work from two research areas, which both provide background for our study. We start with an analysis of existing approaches to assess the programming skills of novice programmers in their early stages of learning. Then we introduce recent work on the application of machine learning for prediction of student performance.

\subsection{Approaches to assess programming skills in early stages of learning}

Lack of progress in the early stages of the learning process can create negative momentum, eventually leading to high failure rates. Alternatively, steady progress can lead to positive momentum as each new concept can help reinforce the earlier foundations.  
Some multi-institutional studies, e.g. \cite{fincher2006predictors}, have used map drawing styles to predict the success of programming among novice programmers, also suggesting that problem solving, and logical thinking are important skills necessary to succeed in the course. Results of many studies have highlighted the need to assess reasoning skills forming the basis for code writing early, through appropriate tasks. 
Some exhaustive descriptive studies, e.g., \cite{simon2012introductory} and \cite{sheard2012exams}.  
  have been conducted to understand the types of questions used in the examinations in programming course. 
 The study of Lopez et al.  
 indicates that there exists a loose hierarchy of skills which the students progress through while learning introductory programming concepts, see \cite{lopez2008relationships}.  
 Activity Diagrams and Parson's Puzzles have been proposed as instruments that to assess the programming ability because they might correlate better with problem solving than code tracing, see \cite{harms2016distractors} and \cite{parsons2015we}. 
 Parson's puzzle tasks were designed to ease novices into code writing by allowing students to piece together code fragments interactively, see~\cite{denny2008evaluating}.  
 The Spearman ranking coefficient for code writing also showed closer correlation with Parson's puzzle questions when compared to tracing. Parson's puzzle questions however, limit students' freedom in arriving at a solution. Multiple-choice questions (MCQs), when appropriately designed, can be really effective for testing intermediate levels of programming skills. MCQs can be easily automated, which is a critical advantage of this assessment type. However, to design  MCQs really well, is not a trivial task. MCQs were found to be the most preferred assessment types in many different domains, see \cite{furnham2011would} and \cite{kuechler2003well}. 
 Students in general felt such tests can improve their exam performance, as they felt more relaxed, see \cite{abreu2018multiple}.

An approach using Activity diagrams (ADs)  was introduced to assess the programming ability of students, see~\cite{parsons2015we}. ADs are used to present visually the logical flow of a computer program, and have a notation for sequence, conditional statements, and loops. The Pearson-product moment correlation between the exam questions and class project mark was used to measure the student ability to write the code. However, this measure may not be accurate as students may collude in outside class activities. 
The importance of developing tasks to help students in their self-assessment was also highlighted in \cite{cutts2019experience}.  
\subsection{Application of machine learning for prediction of students' performance}
\label{sec:relatedML}

Students' performance prediction is one of the earliest and most valuable applications of Educational Data Mining (EDM). 
A systematic literature review (SLR) on the solutions to predict 
student performance using data mining and learning analytics techniques has been presented in \cite{namoun2020predicting}. 
Random Forest (RF) and Linear Regression (LR) algorithms have particularly been used in predicting students' academic performance, where the RF algorithm is in the top 5 algorithms with an accuracy of $98\%$.  
According to \cite{sandoval2018centralized}, 
in an attempt to find a low-case predictive model using the Learning management system data, RF was found to have higher precision in predicting students who are at risk of failing the course i.e., poor performing students.  
A similar result was presented in \cite{chettaoui2021predicting}, where the results of the study demonstrated that, of the five classification algorithms, RF outperformed with $84\%$  accuracy.

The prediction of student performance was applied in the context of e-learning~\cite{abubakar2017prediction}. The study focused on comparing the state-of-the-art classifier algorithms to identify the most suitable for creating a learning support tool. Two comparative studies were conducted using different data sets comprising 354 and 28 records.  Findings of that study revealed that  the Naive Bayes algorithm achieved 72.48\% accuracy, followed by LR with $72.32 \%$.  
To predict student's success in electronics engineering licensure exam, 500 student's data over different cohorts from 2014 - 2019 was used, 
see \cite{maaliw2021early}. The study used RF algorithm to predict the student exam outcome by using 33 different features and had a prediction accuracy of $92.70 \%$.

An extensive evaluation of machine learning algorithms such as Decision Tree, Naive Bayes, Random Forest, PART and Bayes Network was conducted on 412 postgraduate students' data to predict their academic performance in the current semester. The study \cite{kumar2017evaluation} demonstrated that Random Forest (RF) gave the best performance with precision (1) which is essential in identifying students that are likely to fail at the early stages of the course.

In a study presented in \cite{meylani2014predicting}, Neural Network (NN) and LR algorithms were evaluated with respect to prediction of  students' performance in mathematics examination. The LR models (Linear, Multi-nominal and Ordinal)  outperformed the NN models.
 The study used the student's performance to in-class mathematics tests to predict if the students would pass, fail or excel in the final exam.

\section{METHODOLOGY}
\label{sec:methodology}

For the study reported in this paper, data was collected through an in-class test facilitated through Google forms. The test consisted of two parts where the first part included objective questions, and the second part included code writing questions. The first part had 12 objective questions which included tracing questions and three different types of ART questions, explained below. Students were advised to spend about 30 minutes on the objective questions (Tracing and Algorithmic Reasoning Tasks) and the remaining 80 minutes on three code-writing questions.

\subsection{Design of an Abstract Reasoning Task (ART) Type Question}
The ART type questions were designed to get relational level responses from students by making tracing difficult or impossible within allocated time. The ART-Detection question shown in Figure 1 was used in the study, where students had to extract the purpose of the algorithm and express it by writing the correct output for different inputs. Given 6 different input arrays with 7 or more values in each, students who have not extracted the overall effect of the algorithm and apply it to different inputs are unlikely to get all the outputs correct. The students are only awarded the mark if they get all the outputs correct.

\begin{figure}[ht!]
  \centering
{\includegraphics[width=0.45\textwidth]{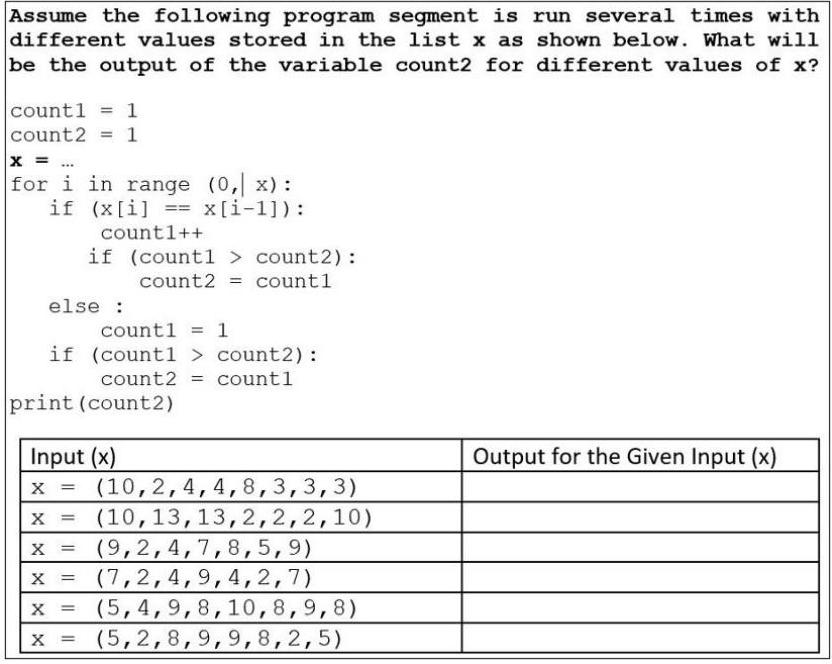}}
\caption{ART Detection Type question used in the study}
  \label{fig:art1}
 \end{figure}

 \begin{figure}[ht!]
\includegraphics[width=0.45\textwidth]{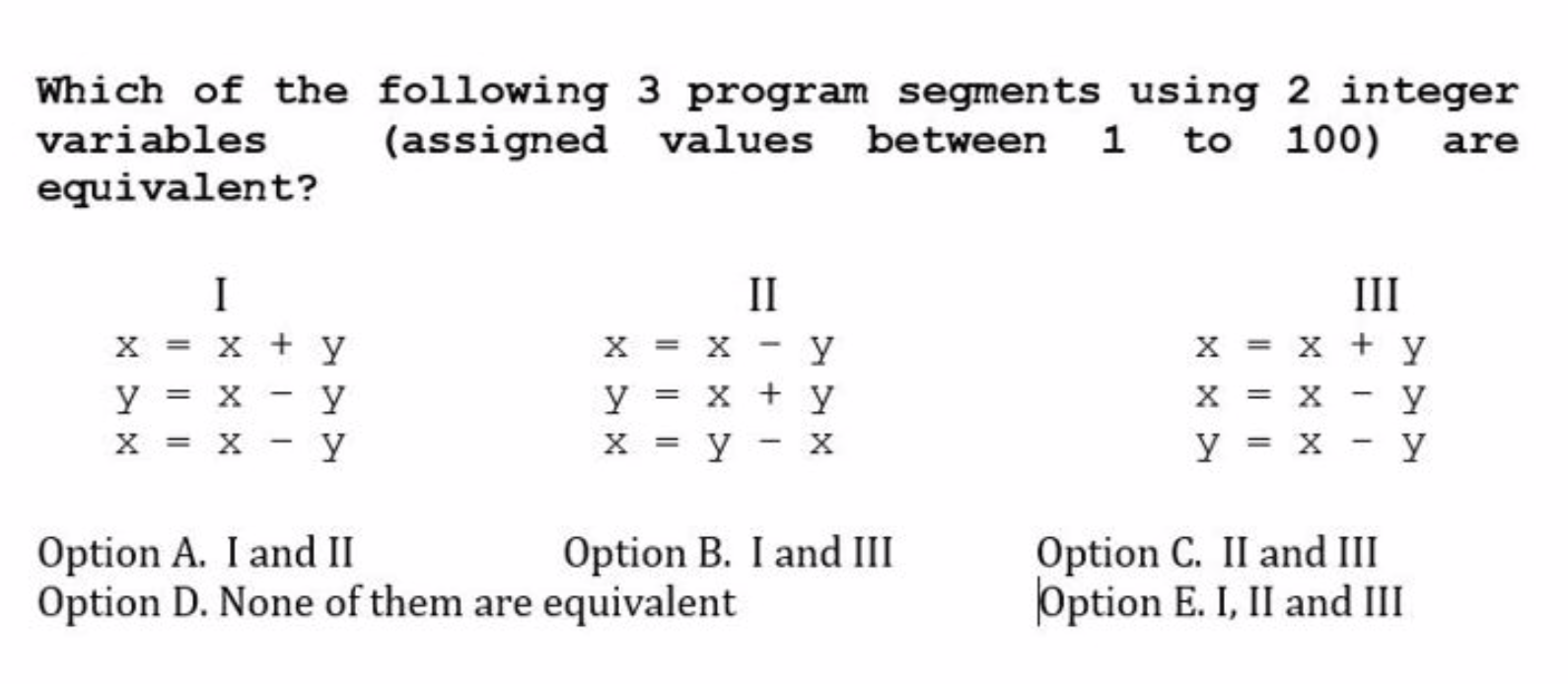}
\caption{ART Comparison Type question used in the study}
  \label{fig:art2}
 \end{figure}

\begin{figure}[ht!]
  \centering 
\includegraphics[width=0.45\textwidth]{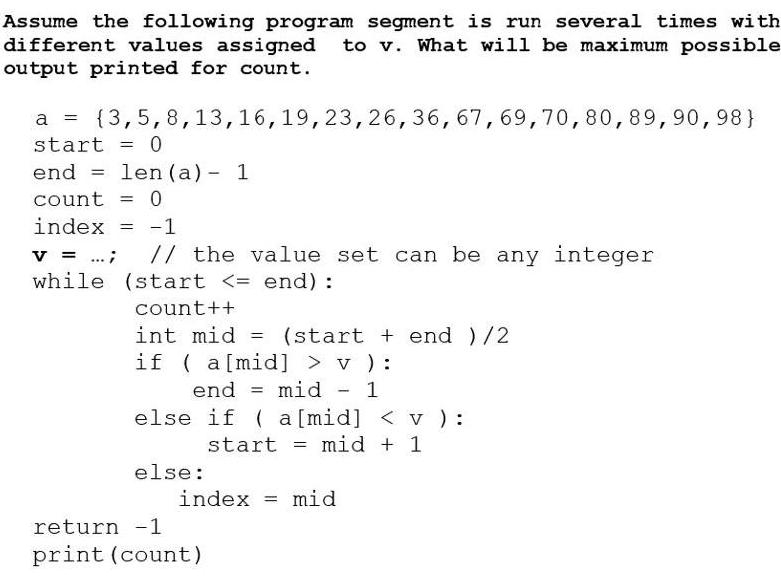}
\caption{ART Analysis Type question used in the study}
  \label{fig:art3}
 \end{figure}

In the past getting relational responses from students primarily relied  on the code-explain instrument. The main drawback of code-explain is the need for manual marking. The ART-detection instrument we have designed requires a similar relational response from students who firstly need to understand the overall effect of the algorithm before applying it repeatedly under the time constraint. Similarly, the ART-comparison instrument requires the students to identify the algorithms that have similar behaviour while the ART-analysis instrument requires students to analyse the algorithm and identify the best- and worst-case scenarios. The SOLO classification of the objective questions used in this study and their purpose are given in the below Table~\ref{tab:table1}. The main advantage with all three ART instruments is that it can be automated and that the marking will not be subjective. The ART-Comparison question and ART-Analysis question used in the study are shown in Figures~\ref{fig:art2} and \ref{fig:art3}.

\subsection{Code Writing}
This test involved three code writing tasks. We consider that the process of writing or creating a block of code to answer the given code writing question involves:

\begin{itemize}
  \item Problem Analysis, which requires analysing and understanding the problem domain;

  \item Solution Planning, which requires creating the steps in coming up with a viable solution (algorithm) to solve 
  the problem;

  \item Coding, which involves converting the problem domain into program domain by combining the different constructs (for-loop or while-loop, etc.) and syntax, to write programming code which computes the required answer(s).
\end{itemize}

\begin{table*}[ht!]
\vspace{-0.2cm}
\caption{Classification of Objective Questions using SOLO Taxonomy
}\label{tab:table1} \centering 
{\footnotesize{ 
\begin{tabular}{|c|c|l|}
\hline
 Question  Type   & SOLO Level & Purpose \\
\hline
Tracing & Multi-Level & 
$\begin{array}{l}  
\text {Requires the students to understand each statement and trace } \\ \text {each line in the in the  given piece code and determine the output}   \end{array}$ \\
\hline
Detection & $\begin{array}{l}\text { Relational } \\ \text { level }\end{array}$ & $\begin{array}{l}\text {Requires abstraction skills to  detect what the role of the algorithm. } \\ \text {Students are expected to apply cognitive  skills at relational level } \\ \text {to analyze how the behavior will  change for different inputs. }\end{array}$ \\
\hline
Comparison & $\begin{array}{l}\text { Relational } \\ \text { level }\end{array}$ & 
$\begin{array}{l}\text {Students are expected to identify algorithms which will display the same } \\ \text {collective or  composite behavior considering different input values. }\end{array}$ \\
\hline
Analysis & $\begin{array}{l}\text { Relational } \\ \text { level }\end{array}$ & $\begin{array}{l}\text {Students are expected to analyze an algorithm including working out } \\ \text {worst case scenarios considering all possible paths. }\end{array}$ \\
\hline
\end{tabular}
} }
\end{table*}

\subsection{Marking}
A positive grading scheme was used for the objective questions: 
\begin{itemize}
    \item 1 mark was awarded for a correct answer,
    \item  0 marks were awarded for an incorrect answer (an answer which is only partially correct was labelled as incorrect), and
    \item no negative marks were awarded for incorrect answers.
\end{itemize}
 For the code writing tasks, partial marking was applied depending on whether the given code was partially correct or fully correct. 
 
 Each Code Writing question carried 3 marks. 
 Grading was done by an experienced lecturer.\\
 ~

\subsection{Choice of Machine Learning Model}

 Because we are predicting the student performance based on their score for an in-class assessment, it is important to ensure the algorithm is outlier resistant and scalable. 
We analysed several automated approaches to predict the students' performance, see Section~\ref{sec:relatedML}. Their results highlighted that application of the Random Forest algorithm is the most promising direction for prediction of students' performance.  
We trained the model using both Random Forest (RF) and Ordinal Logistic Regression (LR) algorithms because of the inherent nature of the variables. Therefore, in this this study we report the performance of both these models in predicting the student performance in code writing.  

\subsection{Data Exploration and Training}
In this study we have used the students' performance on the objective questions (tracing and ART questions) to predict students' success in code writing. Data used in this study consists of 243 students' performances on 15 programming questions including both objective and code writing questions. The students' data was loaded into Google drive and mounted on to Google Colab Notebook.  

\subsubsection{Data Preparation} 
We performed the data profiling using the python panda library to visualise the data. Data pre-processing was performed to identify any duplicates, missing values and data types. The data was encrypted to hide any student's identifying information to ensure privacy and confidentiality. The data encoding of the independent variables was performed to ensure all the data in the dataset are of the same datatype. 
The unique code writing marks in the dataset $0,0.5,1,1.5,2,2.5$ and 3 were encoded to $0,1,2,3,4,5$ and 6 respectively.

The next step in the data preparation was to separate the dependent and independent variables:
\begin{itemize}
    \item 
    The \emph{dependent variables} are the features which we train the model on. In our study, the features are the students' scores to the objective questions. 
    \item 
    The \emph{independent variables} are also called the labels. 
    In our study, these are the students score to code writing programming questions.
\end{itemize}
 From our data exploration using the data profiling in Python we found that the data set had a class imbalance. To overcome the issue of class imbalance in the dataset the following steps were performed:

\begin{itemize}
  \item Step 1: Firstly, we identified the class labels $(0,1,2,3,4,5,6)$ whose frequency in the dataset is less than 10 and these class labels were removed from the dataset.

  \item Step 2: The dataset is then split into training and testing dataset using the \emph{train\_test\_split\_function} from Scikit-learn (sklearn) model selection.

  \item Step 3: The class labels whose frequencies were less than 10 (identified in Step 1) were added to the training split of the dataset to ensure that the machine learning algorithm had enough data on these labels to train on. The dataset is now ready to be trained by the chosen machine learning algorithms. 

\end{itemize}

\subsubsection{Data Training and Performance Metrics}
The training split of the dataset was then trained using the Scikit-learn (sklearn) ensemble \emph{RandomForestregresso} and \emph{LIBLINEAR} Python library. The 10-fold (k-fold) cross validation technique was then applied to both the models using the \emph{GridSearchCV} function from sklearn. 10-fold cross validation involves randomly dividing the training data into 10 folds. The first fold was considered as the testing set and the model was fitted on the remaining 9 folds of the data.

\subsubsection{Performance Metrics} 
We have compared the effectiveness of both the trained data models using the following metrics shown in below Table~\ref{tab:table2}.  In addition to the above metrics, different stratified  
training and test splits were used to evaluate the performance of the models. 

\begin{table*}[ht!]
\vspace{-0.2cm}
\caption{Performance metrics to evaluate the \emph{modelPerformance} metrics to evaluate the model}\label{tab:table2} \centering
{\footnotesize{
\begin{tabular}{|l|l|l|}
\hline
Metric & Definition & Formula \\
\hline
$Accuracy$ & 
$\begin{array}{l}\text { It is the proportion 
of correction predictions, i.e.,  } \\ \text {both true positives (TP) and true negatives (TN), } \\ \text {among total number of cases examined }\end{array}$ & $=(\mathrm{TP}+\mathrm{TN}) /(\mathrm{TP}+\mathrm{TN}+\mathrm{FP}+\mathrm{FN})$ \\
\hline
$Recall$ & 
$\begin{array}{l}\text {The ability of a model  to find all the relevant } \\ \text { cases within a data set }\end{array}$ & $=(\mathrm{TP} /(\mathrm{TP}+\mathrm{FN}))$ \\
\hline
$Precision$ & 
$\begin{array}{l}\text { Precision, also known  as positive predictive  value, } \\ \text {is the fraction  of relevant instances  among the } \\ \text { retrieved instances }\end{array}$ & $=(\mathrm{TN} /(\mathrm{TN}+\mathrm{FP}))$ \\
\hline
$F-Measure$ & 
$\begin{array}{l}\text {It is the harmonic  mean of precision and   recall }\end{array}$ &  $=(2*Precision*Recall)/(Precision +Recall) $\\
\hline
\end{tabular}}}
\end{table*}

\section{RESULTS}
\label{sec:results}
Table~\ref{tab:table3} presents a performance comparison of the Random Forest (RF) and Logistic Regression (LR) models for different training and test splits in predicting students' performance to code writing questions. RF attained the highest cross-validated (10-fold) accuracy score of $85.45 \%$. In comparison to the  LR model, the RF model has a high precision of 
predicting students who are at the risk of failure (getting 0 or 0.5 marks in code writing).

\begin{figure}[ht!]
  \centering
\includegraphics[width=0.45\textwidth]{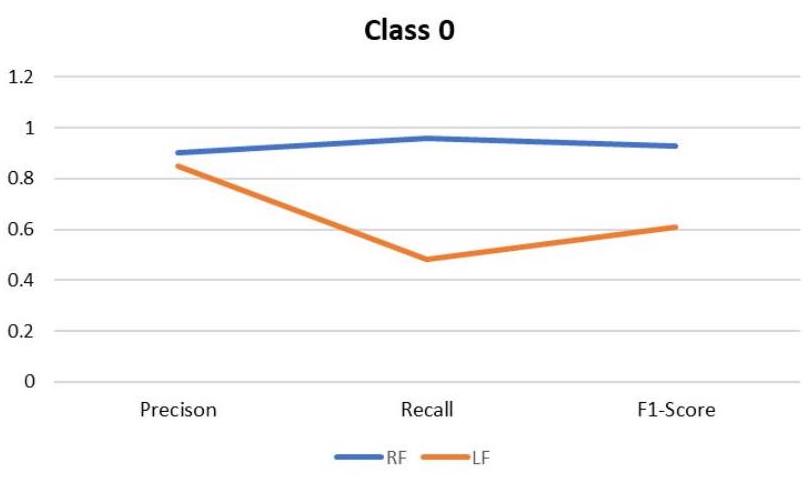}
  \caption{Performance metrics of RF and LF models in predicting the students scoring 0 marks}
  \label{fig:fig4-performance0}
 \end{figure}

\begin{figure}[ht!]
  \centering 
\includegraphics[width=0.45\textwidth]{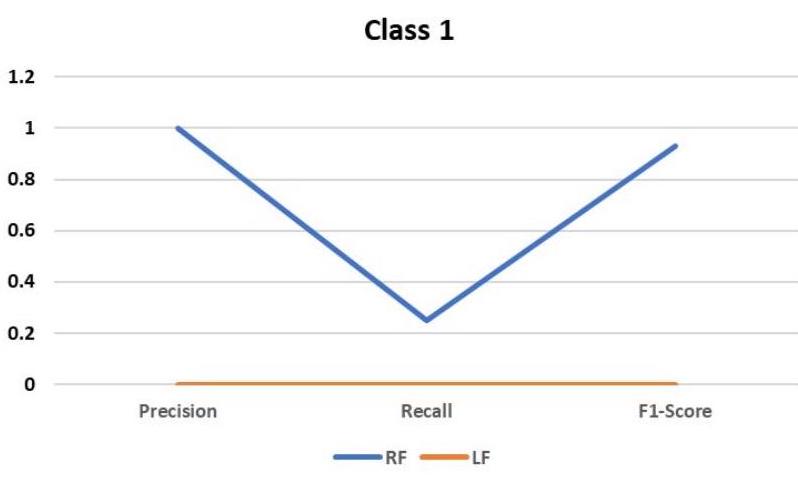}
  \caption{Performance metrics of RF and LF models in predicting the students scoring 0.5 marks}
  \label{fig:fig5-performance05}
 \end{figure}

Figures~\ref{fig:fig4-performance0} and \ref{fig:fig5-performance05} demonstrate the precision, recall and F1-score of both models in predicting students that are at risk of failure. According to our analysis, RF algorithm performs better compared  
to LR in precision, recall and F1-measure: RF has a high precision of 0.90 and 1 in predicting students that are to score 0 and 0.5 marks in code writing tasks.   
This precision is vital in identifying students that are at risk to fail in the programming course. Based on our analysis it was clear that RF was more suitable models for predicting the student performance, which also correlated with the results presented in the related works, see Section~\ref{sec:relatedML}.

To further evaluate the RF model for applying the ART type questions to predict the students code writing performance,  we have analysed the \emph{feature importance}: 
The RF model computes which feature contributes the most to decrease the weighted impurity. The feature with highest value contributes the most in predicting the variable i.e, code writing. 
We have used the Scikit-learn to obtain the feature importance. 

\begin{figure}[ht!]
  \centering 
\includegraphics[width=0.45\textwidth]{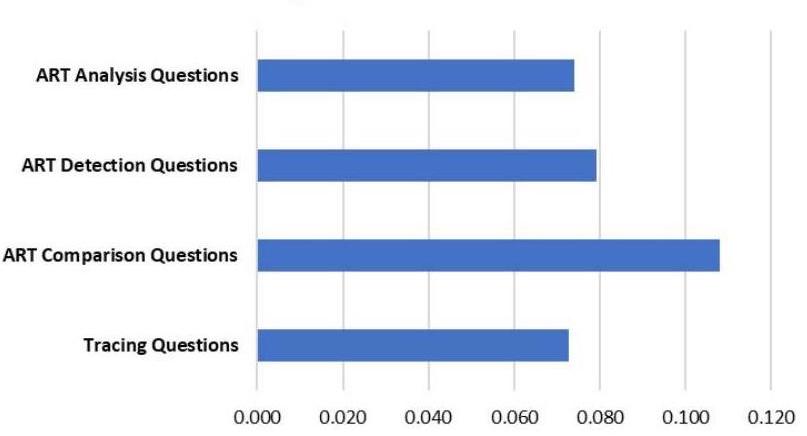}
  \caption{Feature Importance of Random Forest}
  \label{fig:fig6-imp}
 \end{figure}

Figure~\ref{fig:fig6-imp} shows the feature importance computed by the Random Forest model. Based on our analysis, we found that the student's performance on ART Comparison type questions (0.11) had a significant influence on the student code writing results, followed by ART Detection type questions (0.079) and ART Analysis questions (0.074). The least contribution was from the tracing question grades (0.072). This confirms that the RF model uses the ART type questions to predict the students code writing abilities. 

\begin{table}[h]
 \vspace{0.4cm}
\caption{Comparison of accuracy  of RF and LF models}\label{tab:table3} \centering
\footnotesize{\begin{tabular}{|l|l|l|}
\hline 
Train-Test split & \multicolumn{2}{c|}{Accuracy}\\
~  & RF & LR \\
\hline
$75-25$ & $85.45 \%$ & $46.45 \%$ \\
\hline
$70-30$ & $81.82 \%$ & $41.81 \%$ \\
\hline
\end{tabular} }
\end{table}

\begin{table}[ht!]
\vspace{0.4cm}
\caption{Spearman Rank Correlation with Code Writing for tracing, comparison ($C$), detection ($D$), analysis ($A$) and average correlation for all ART instruments}\label{tab:table4} \centering
\footnotesize{
\begin{tabular}{|c|c|c|c|c|}
\hline
Non-ARTs & \multicolumn{4}{|c|}{ARTs} \\
\hline
Tracing & $C$  & $D$ & $A$ & average \\
\hline
0.63 & 0.69 & 0.68 & 0.74 & 0.70 \\
\hline
\end{tabular}}
\end{table}

\begin{table}[ht!]
\vspace{0.4cm}
\caption{Comparison of Pearson correlations of ART, Activity Diagrams, and Parson's Puzzle with code writing (the data on both Activity Diagrams and Parson's Puzzle have been provided in~\cite{parsons2015we}) }
\label{tab:table5} \centering
\footnotesize{
\begin{tabular}{|c|c|c|}
\hline
ART & Activity Diagrams & Parson's Puzzle \\
\hline
0.37 & $0.26$ & $0.12$ \\
\hline
\end{tabular}}
\end{table}

\newpage 
\noindent
The Spearman correlation coefficients, presented in Table~\ref{tab:table4}, affirm  a very strong positive relationship between ART Analysis type questions and the code writing question at 0.74. In addition, ART Comparison (0.69) and ART Detection (0.68) have a positive correlation with code writing. 
The ART type instruments as a whole have higher positive correlations of 0.70 compared to what we can get for tracing (non-ART instrument), which was only 0.63.

Based on our findings summarized in Tables~\ref{tab:table3} and \ref{tab:table4}
as well as Figure~\ref{fig:fig6-imp}, it is clear that the ART instrument types have strong connection to students' code writing performance. Thereby confirming ART instruments type can assess the needed relational skills required to excel in coding. This confirms that ART instruments can be used to assess the programming abilities of the students and to identify the students who are at the risk of failure in the early stages of the course. 
Moreover, comparing our results with the results on Activity Diagrams and Parson's Puzzles from \cite{parsons2015we},  we  found that the ART instruments have a slightly higher Pearson correlation (0.37) compared to Activity Diagrams (0.26) and Parson's Puzzles (0.12), see Table~\ref{tab:table5}.

Based on the above findings, our study affirms that ART type instruments can be used to assess the students' programming abilities at the early stages of the course to reduce failure rates and the RF model is more suitable compared to LR to automatically identify students who might require additional support for learning code writing skills.

\section{CONCLUSION AND FUTURE WORK}
\label{sec:conclusion}

In many cases, students  struggle with programming mainly through the lack of problem-solving ability. Problem solving ability is also central for students having to apply scientific and mathematical principles to solve real world problems. In recent years, learning trajectories resulting from a research-based curriculum development approach have benefited student learning mathematics, by modelling their thinking process, see \cite{izu2019fostering}. However, there has been little research done on how such an approach can foster problem solving and code writing skills in novice programmers.

Our research has therefore focused on supplementing traditional and well-researched instruments such as program tracing with tasks demanding more relational thinking gradually. We have developed ART-type questions such as comparison, detection, and analysis which require students to map from problem domain to the solution domain (coding). We are following a scientific approach by using machine learning models that identify tasks which are progressively more complex but lead gradually to skills needed for problem solving and code writing. The main novelty of our approach is the ability to automate the process of assessment feedback by creating a trajectory of tasks which require no manual intervention and by predicting the students programming abilities based on these trajectory tasks. 

The data collected with students have clearly revealed tasks demanding relational level responses which better correlate with code writing when compared to tracing. 
Our approach involved developing a trajectory of tasks rooted in a multidimensional framework that combined different levels of the SOLO taxonomy with multiple domains. It is our belief such an approach can lead to better learning outcomes in coding as code writing requires analytical ability to understand the problem domain, as well as abstraction skills and the ability to come up with algorithms which can be coded and implemented.

ART type instruments designed to get relational level responses showed greater correlation with code writing when compared to tracing. The Random Forest regression model had an accuracy of $84.5 \%$ in predicting student success in code writing based on algorithmic reasoning tasks. The Spearman Rank Correlation coefficient was substantially higher for ART types when compared to tracing. Within the Algorithmic Reasoning Tasks, comparison questions showed substantially higher feature importance when compared to tracing, ART-detection and ART-analysis. Our preliminary results show new types of instruments that gather relational responses can be developed resulting in greater similarity to the reasoning skills needed in code writing. We have shown that these ART type instruments form a loose trajectory starting from Code Tracing Questions to ART Analysis questions, to ART Detection questions, to ART Comparison questions to code writing. By classifying such instruments using machine learning, novices can be provided with a learning trajectory that equips them better with the cognitive skills needed for code writing.

Our preliminary studies with tasks combining multiple domains (problem domain and coding) suggest these tasks demand even greater cognitive depth. To further evaluate we would consider collecting more data over the next few semesters. We also aim to use the final exam results and predict the student success in passing or failing the exam based on how well they perform in ART questions in their mid-semester. We also would like to identify more instruments that can be used to assess the students' programming abilities.

\bibliographystyle{apalike}

\end{document}